\documentclass[twocolumn,superscriptaddress]{revtex4}
\usepackage{color}
\usepackage{graphics,graphicx,epsfig}
\usepackage{epsf,epstopdf,wrapfig}
\usepackage{amssymb,amsfonts,amsmath}
\include{epsf}
\usepackage{ifthen}		
\usepackage{url}

\newcommand{\beqn}{\begin{eqnarray}}
\newcommand{\eeqn}{\end{eqnarray}}
\newcommand{\beq}{\begin{equation}}
\newcommand{\eeq}{\end{equation}}

\usepackage{ifthen}

\begin{document}


\newcommand{\format}{pnasfigintext}  

\newcommand{\onlycaption}{true}

\newcommand{\panelone}{
\begin{figure}
\begin{center}
\ifthenelse{\equal{\onlycaption}{false}}{}{
\noindent\includegraphics[width=\linewidth]{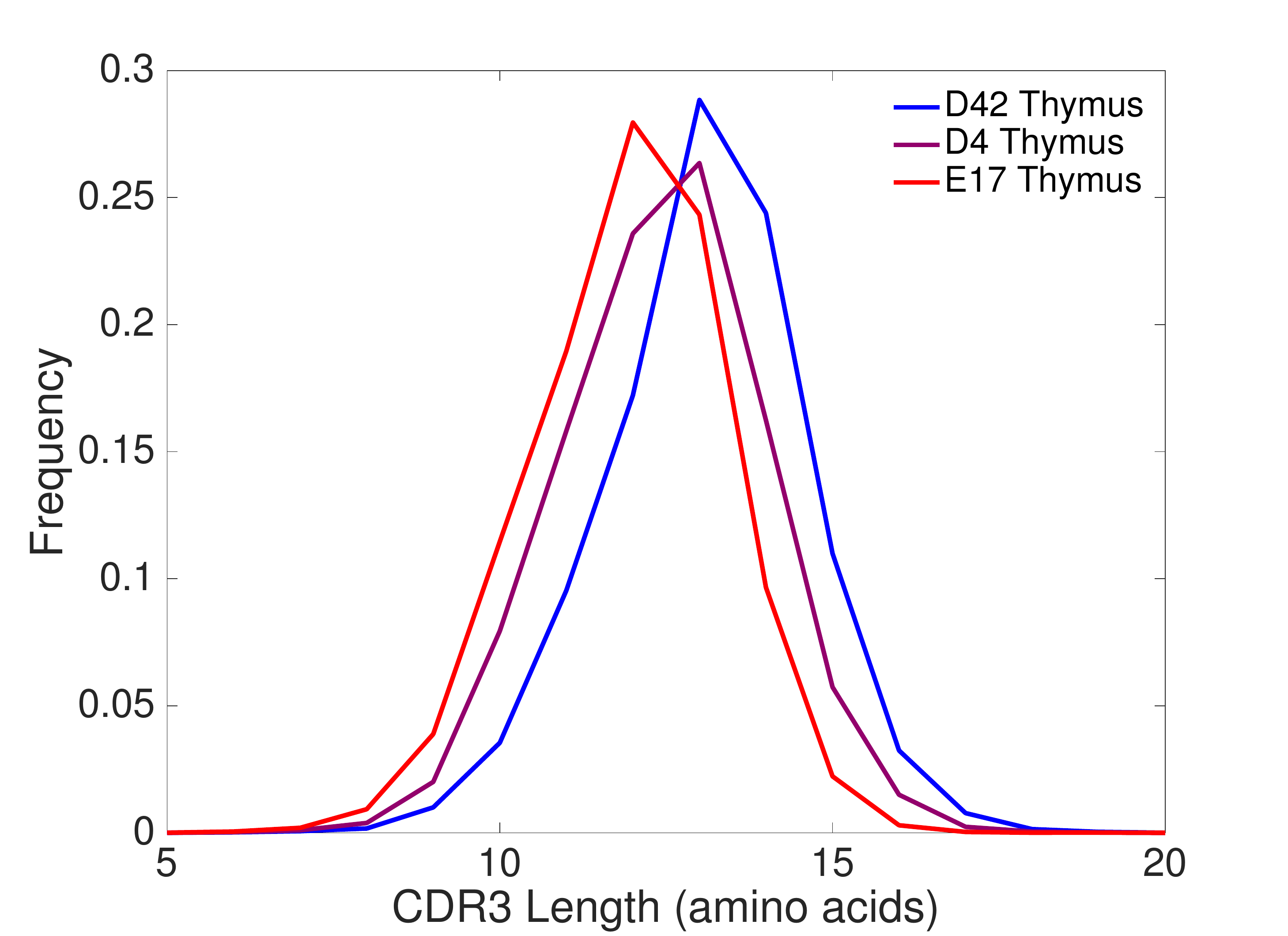}
}
\caption{{\bf Distributions of CDR3$\beta$ amino-acid length at different ages.} Data derived from mouse thymus at three ages: 17 day embryo (E17), 4 days (D4) and 42 days (D42) post birth. In passing from embryo to mature animal, the mean CDR3 length increases by just one amino acid. \label{fig1}
}
\end{center}
\end{figure}
}

\newcommand{\paneltwo}{
\begin{figure}
\begin{center}
\ifthenelse{\equal{\onlycaption}{false}}{}{
\noindent\includegraphics[width=\linewidth]{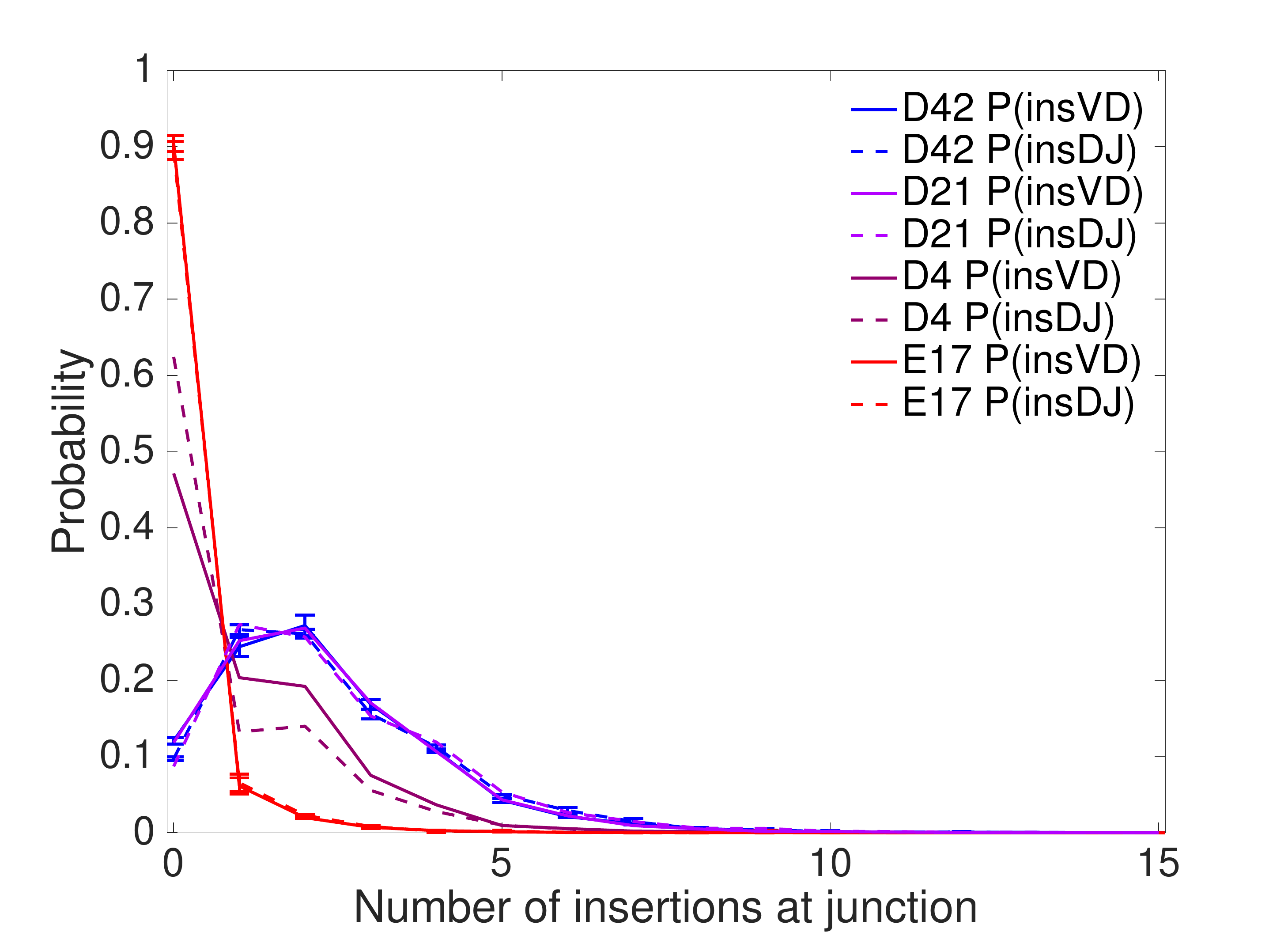}
}
\caption{{\bf Age-dependent insertion length distributions.} Distributions of the number of N insertions at the VD junction, $P({\rm insVD})$, and at the DJ junction, $P({\rm insDJ})$, inferred from individual mouse thymus data sets at different ages: embryo day 17  (E17), and 4, 21 and 42 days post birth (D4, D21, and D42). The error bars indicate the variation across individuals (see SI Table 1). \label{fig2}
}
\end{center}
\end{figure}
}

\newcommand{\panelthree}{
\begin{figure}
\begin{center}
\ifthenelse{\equal{\onlycaption}{false}}{}{
\noindent\includegraphics[width=\linewidth]{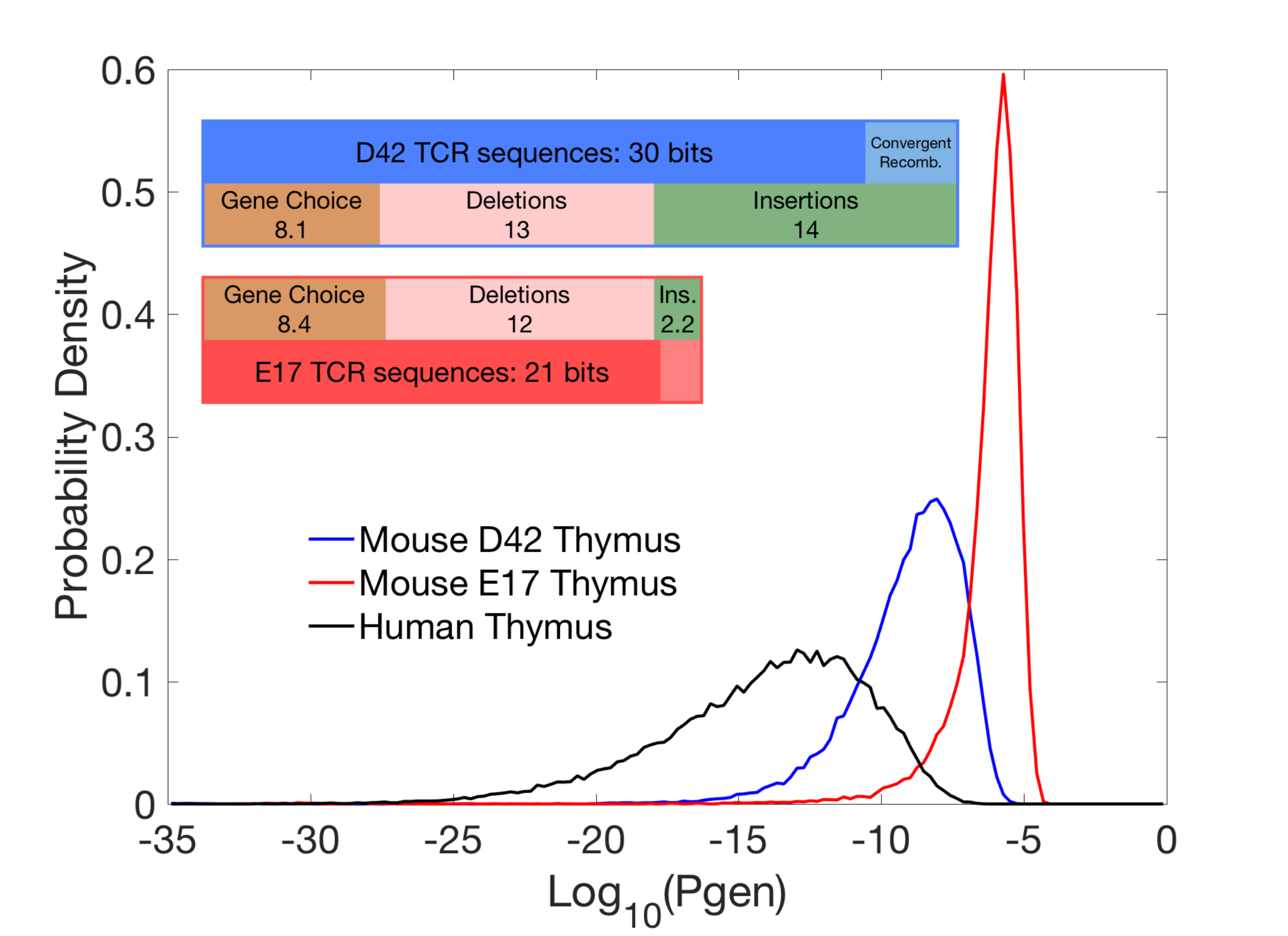}

}
\caption{ {\bf Sequence entropy for thymic repertoires.}  Distribution of the log generation  probability for mouse thymic repertoires derived from 17 day embryo (E17) and 42 day post birth animals (D42). The human generative probability distribution \cite{Murugan2012} is plotted for comparison. Shannon entropy is minus the mean over these distributions. Inset: Shannon entropy at ages 17 days embryo and 42 days post-birth, decomposed into the components of the recombination scenario. Sequence entropy is recombination scenario entropy minus a correction for convergent recombination. Note that only the insertion component of scenario entropy changes significantly between embryonic and mature.\label{fig3}
}
\end{center}
\end{figure}
}

\newcommand{\panelfour}{
\begin{figure}
\begin{center}
\ifthenelse{\equal{\onlycaption}{false}}{}{
\noindent\includegraphics[width=.8\linewidth]{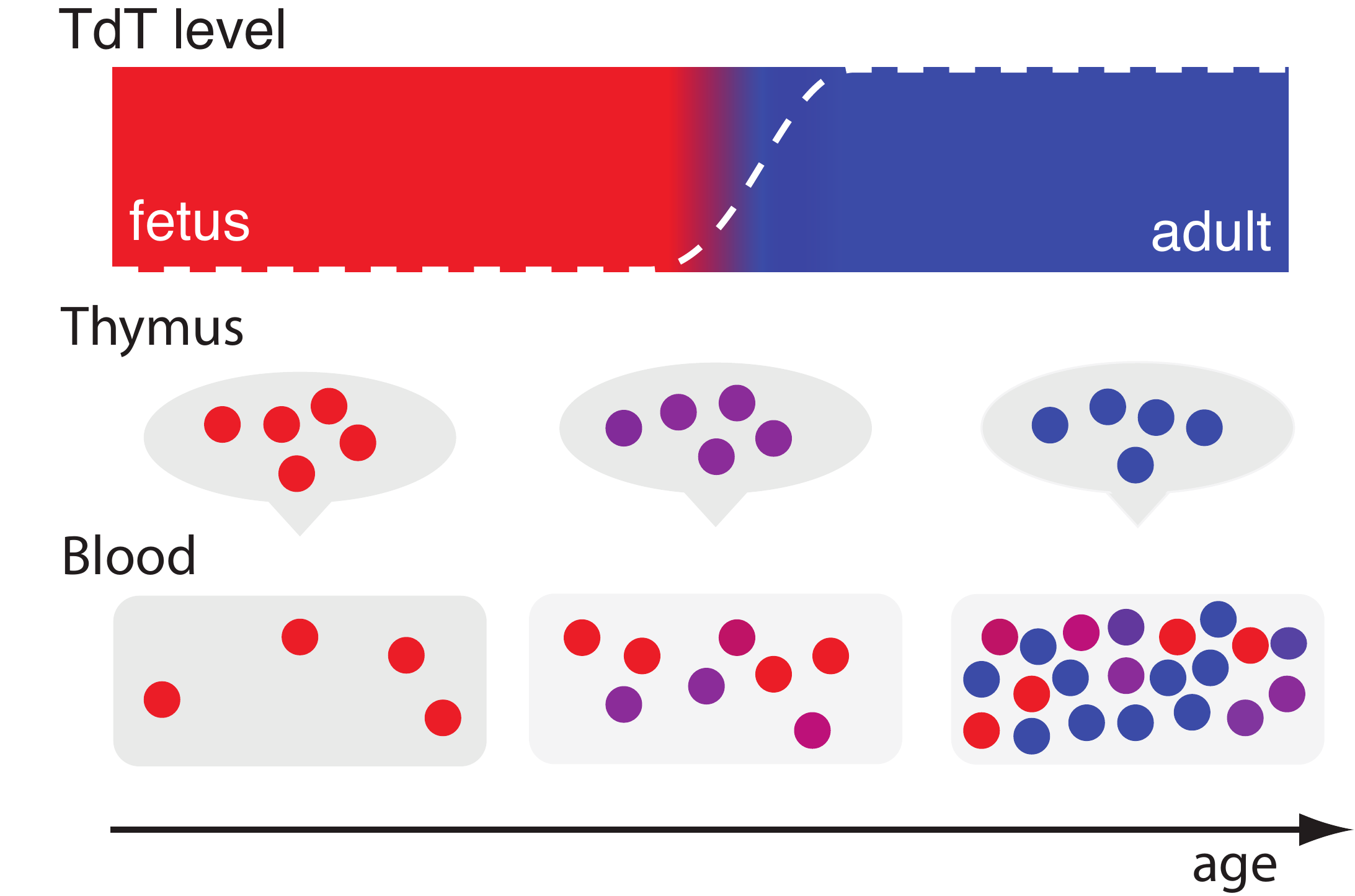}
}
\caption{ {\bf Mixture model dynamics.} T cells recombine according to an insertion profile that depends on a time-dependent effective TdT level $\alpha$ (dashed white line in upper panel). Cells output to the periphery at different times are described by different values of $\alpha$ (indicated by colors). The accumulating peripheral T cell repertoire is described by a mixture that accounts for the different numbers of T cells emitted at different times. \label{fig4}
}
\end{center}
\end{figure}
}

\newcommand{\panelfive}{
\begin{figure}
\ifthenelse{\equal{\onlycaption}{false}}{}{
\noindent\includegraphics[width=\linewidth]{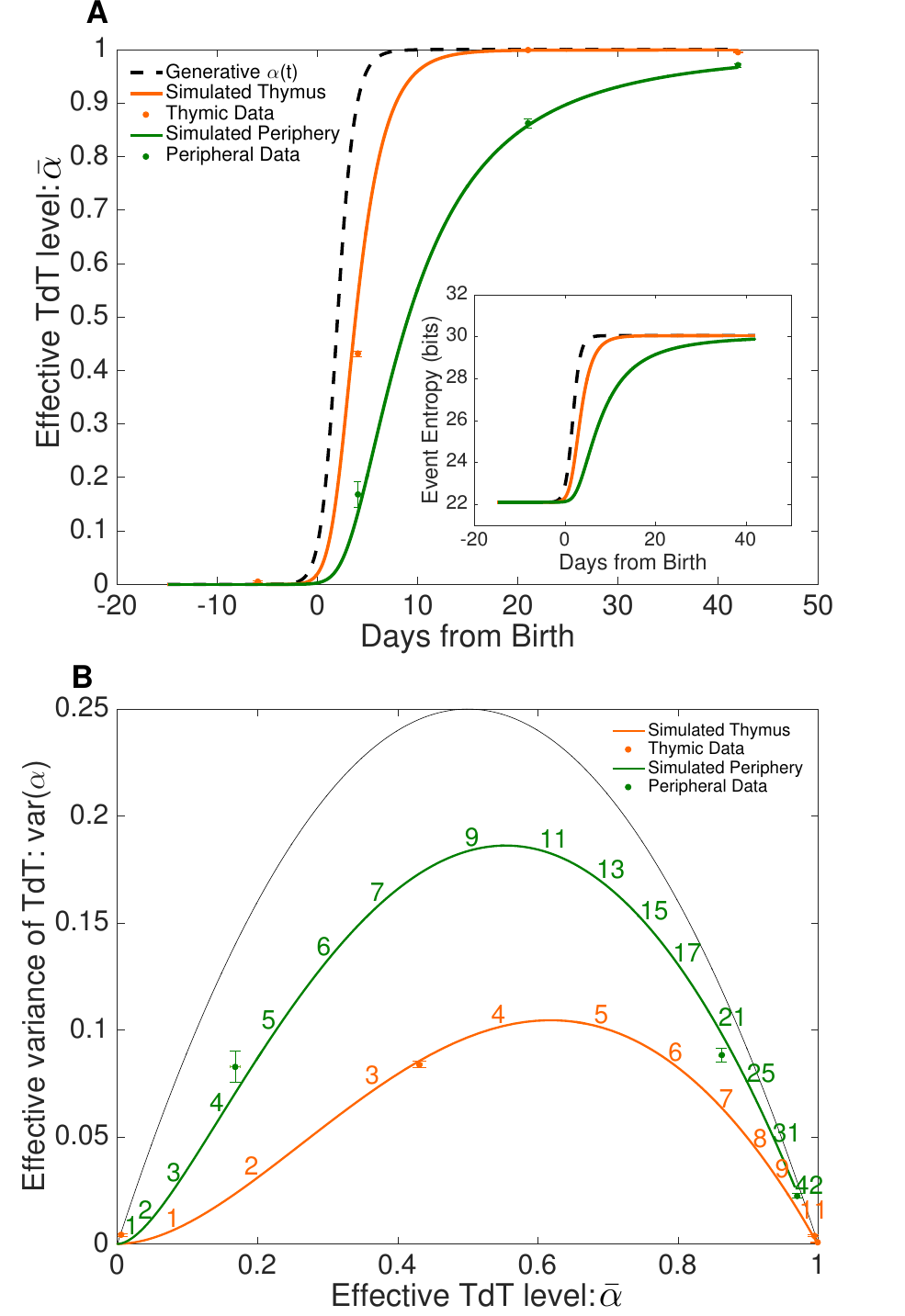}
}
\caption{{\bf Age dependence of the recombination process.} The effective TdT level $0\leq \alpha\leq 1$ is estimated as an interpolation parameter between the recombination statistics of embryonic and mature animals.
A) Mean effective TdT $\bar\alpha$ at various ages (17 day embryo and 4, 21, and 42 days post-birth), from different tissues: thymus and periphery. Periphery is taken from blood, except for day 4 for which it is taken from spleen. The data points are compared to a minimal model of thymic entry, residence and output, with a sharply increasing effective TdT level $\alpha(t)$ represented by the dashed line (see Materials and Methods). Inset: recombination entropy as a function of age, as predicted by the model.
B) The variance of $\alpha$, which measures the level of mixing in the repertoires, is shown as a function of its mean for both data (symbols) and the prediction of the minimal model (lines). The black line shows the maximal possible variance $\bar \alpha (1-\bar\alpha)$. Numbers represent age from birth in days.
\label{fig5}
}

\end{figure}
}

\newcommand{\panelsix}{
\begin{figure}
\ifthenelse{\equal{\onlycaption}{false}}{}{
\noindent\includegraphics[width=\linewidth]{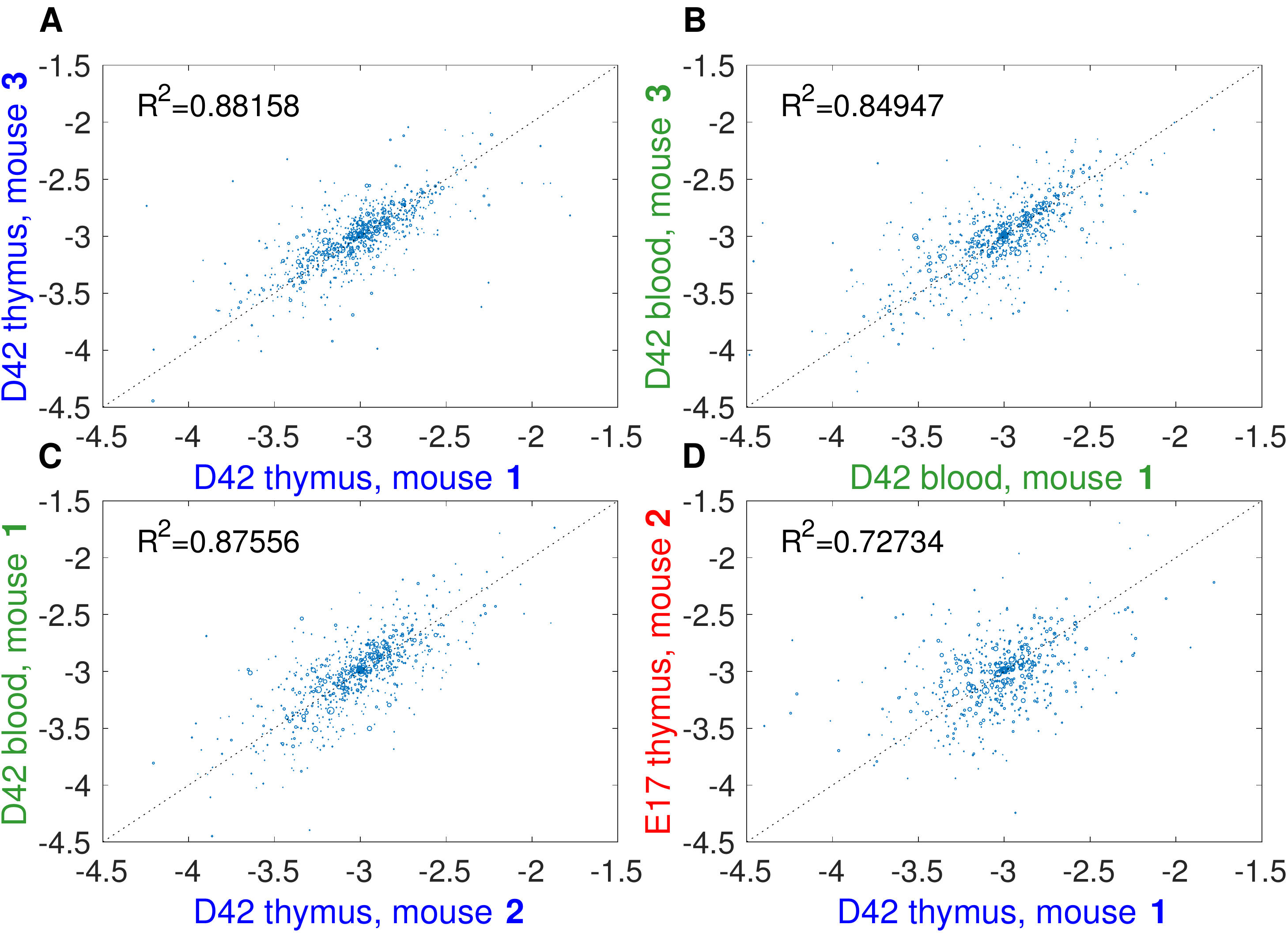}
}
\caption{{\bf Comparing selection factors.}
Scatter plots of the of the logarithms of the $q_{i;L}(a)$ selection factors 
inferred from different pairs of samples. Each dot corresponds to an $(a,i,L)$ combination, and its size is set by the frequency of this combination (frequencies below $.001$ not shown). $R^2=1-\textrm{Var}(q'-q)/(\textrm{Var}(q')+\textrm{Var}(q))$, where $q$ and $q'$ are the selection factors in each sample, and where variances are weighted by frequency. The compared samples are: A) thymus of two mature mice; B) blood of two mature mice; C) blood and thymus of two mature mice; D) thymus of an embryo and a mature mice.
\label{fig6}
}
\end{figure}
}

\title{Mouse T cell repertoires as statistical ensembles: overall
  characterization and age dependence}

\author{Zachary Sethna}
\thanks{These two authors contributed equally}
\affiliation{Joseph Henry Laboratories, Princeton University,
  Princeton, New Jersey 08544 USA}
\author{Yuval Elhanati}
\thanks{These two authors contributed equally}
\affiliation{Laboratoire de physique th\'eorique, UMR8549, CNRS and \'Ecole normale sup\'erieure, 24, rue Lhomond, 75005 Paris, France}
\author{Crissy S. Dudgeon}
\affiliation{Cancer Institute of New Jersey, New Brunswick, New Jersey
  08903 USA}
\author{Curtis G. Callan Jr.}
\affiliation{Joseph Henry Laboratories, Princeton University, Princeton, New Jersey 08544 USA}
\author{Arnold Levine}
\affiliation{Institute for Advanced Study, Princeton New Jersey 08540 USA}
\author{Thierry Mora}
\affiliation{Laboratoire de physique statistique, UMR8550, CNRS and \'Ecole normale sup\'erieure, 24, rue Lhomond, 75005 Paris, France}
\author{Aleksandra M. Walczak}
\affiliation{Laboratoire de physique th\'eorique, UMR8549, CNRS and \'Ecole normale sup\'erieure, 24, rue Lhomond, 75005 Paris, France}

\date{\today}

\begin{abstract}
The ability of the adaptive immune system to respond to arbitrary pathogens stems from the broad diversity of immune cell surface receptors (TCRs). This diversity originates in a stochastic DNA editing process (VDJ recombination) that acts each time a new immune cell is created from a stem cell. By analyzing T cell sequence repertoires taken from the blood and thymus of mice of different ages, we quantify the significant changes in this process that occur in development from embryo to young adult. We find a rapid increase with age in the number of random insertions in the VDJ recombination process, leading to a dramatic increase in diversity. Since the blood accumulates thymic output over time, blood repertoires are mixtures of different statistical recombination processes and, by unraveling the mixture statistics, we can obtain a clear picture of the time evolution of the early immune system.  Sequence repertoire analysis also allows us to detect the effect of selection on the output of the VDJ recombination process. The effects we find are nearly identical between thymus and blood, suggesting that they mainly reflect selection for proper folding of the TCR receptor protein. 
\end{abstract}

\maketitle


The adaptive immune system relies on the diversity of its receptors to recognize an unpredictable array of foreign pathogens. This diversity can now be studied in great detail thanks to high-throughput sequencing of lymphocyte receptor repertoires \cite{Weinstein2009, Boyd2009, Robins2009,Benichou2012,Baum2012,Six2013,Georgiou2014,Calis2014}. Repertoire sequence data can be used to better understand how immune system diversity develops and matures over a lifetime. While the decline in immune diversity with advanced age has been well studied \cite{Goronzy2005,Allman2005b,Goronzy2007,Dunn-Walters2010}, including recently by sequencing T-cell repertoires \cite{Qi2014,Britanova2014a,Britanova2016,Pogorelyy2016}, the development of receptor diversity in early life remains to be fully characterized.
The flexibility of the mouse model makes it an excellent experimental platform for studying the rapid changes in the mammalian immune system that occur as an individual passes from embryo to young adult. In this paper, we use sequence data on T cells collected from mice at several times in early life to study this maturation process. 

T cells perform their surveillance function through surface receptor proteins, or T-cell receptors (TCR). These receptors are highly diverse: any individual TCR recognizes only a small subset of the possible foreign peptides that can be displayed on the surface of somatic cells, and the system defends against a broad range of pathogens by having a very diverse repertoire of T cells. This diversity is achieved by a process of stochastic DNA editing, known as VDJ recombination, which occurs in each newly-produced T cell, and for each of the two chains (alpha and beta) that comprise the receptor. The beta chain, which is the main focus of this paper, is composed of V, D, and J segments, each randomly chosen from a number of  variants linearly arrayed on the germline genome and joined together. Further diversity is created by random deletions from each end of the joined genomic segments, and from random non-templated insertions between them.

The quantitative details of how VDJ recombination operates {are} known to change during development. For instance, B- and T-cell receptors formed in embryonic or neonatal individuals are less diverse than in adults, due to a lower number of non-templated insertions \cite{Feeney1990}, caused by the absence of Terminal Deoxynucleotidyl Transferase (TdT) expression, the enzyme responsible for these insertions \cite{Bogue1992,Komori1993a,Gilfillan1993a}. While this observation has been confirmed by deep sequencing of human TCR repertoires \cite{Britanova2016,Pogorelyy2016}, the precise form and time-resolved dynamics of the maturation of the VDJ recombination process has not been assessed. A direct way to observe the effects of this maturation is to track the length distribution of the Complementarity Determining Region 3 (CDR3) of the beta chain, which streches from a conserved cysteine in the V segment to a conserved phenylalanine in the J segment, and whose length is crucially affected by the addition of non-templated nucleotides. Fig.\ \ref{fig1} shows the evolution of the CDR3 length distribution for T cells extracted from mouse thymus at ages ranging from 17 days after conception to 42 days post-birth (see below and Materials and Methods for details).  The mean CDR3 length increases systematically with developmental age, consistent with increasing numbers of non-templated insertions, meaning that the apparatus that generates the immune repertoire changes with time in just days.

Changes in CDR3 length distribution reflect all the changes that may occur in the statistics of the recombination process, including the distribution of insertions, deletions, and choice of gene templates. However, extracting these distributions directly from sequence data using standard assignment tools \cite{Brochet2008,Thomas2013,Bolotin2015,Yu2015} may give biased estimates that would limit our ability to detect the developmental changes that interest us here.
In previous work we developed a method to overcome this problem using a probabilistic approach \cite{Murugan2012,Elhanati2016}. Here we apply this method to thymic and peripheral TCR beta (TRB) repertoires of mice of varying ages from prenatal to young adult. Our modeling approach allows us to quantify and track diversity as a function of developmental age, in both the thymus and the periphery.

\panelone

\section{Results}

\subsection{Inferring the statistics of VDJ recombination of murine TCR beta chains}
A new TRB gene is created from germline DNA by a series of stochastic events: choosing gene segments, deleting bases from the ends of the chosen gene segments, and inserting nucleotides between the modified gene segments.
Because the same sequence can be generated by distinct recombination events,
these specific events are unknown and cannot be read off from the sequence with certainty. However, the distribution of these events can be inferred from large numbers of sequences, using probabilistic modeling.

We assume that the probability distribution for the generative events involved in VDJ recombination of TRB has the following form \cite{Murugan2012,Elhanati2015}, which we will verify {\em a posteriori}:
\beq\label{eq:pgen}
\begin{split}
P_{\rm }(S)=&P({\rm V})P({\rm D,J})P({\rm insVD},{\rm insDJ})\\
&\ P({\rm delV}|{\rm V})P({\rm delDl},{\rm delDr}|{\rm D})P({\rm delJ}|{\rm J})\\
&\ P(s_1)P(s_2|s_1)\cdots P(s_{\rm insVD}|s_{\rm insVD-1})\\
&\ P(t_1)P(t_2|t_1)\cdots P(t_{\rm insDJ}|t_{\rm insDJ-1})
\end{split}
\eeq
where $S$ is a recombination scenario (defined by gene choice, numbers of deletions, number and identity of insertions) and where each factor in the equation is a distribution over the possible elements of the scenario: $P({\rm V})P({\rm D,J})$ is the distribution of choices of the three kinds of gene segments (note that a correlation between the two D genes and the two clusters of J genes is imposed by genome topology), $P({\rm delV}|{\rm V})$ is the distribution of numbers of deletions from the end of a particular gene V (and likewise for D and J). A scenario includes specific N nucleotide insertions $s_1 ... s_{\rm insVD}$ and $t_1 ... t_{\rm insVD}$ at the VD and DJ junctions, and $P({\rm insVD}, {\rm insDJ})$ is the distribution of the total numbers of such insertions, while $P(s_i|s_{i-1})$ etc. describes the probability of inserting particular N nucleotides.
Note that Eq.\ \ref{eq:pgen} gives the probability of recombination scenarios, not sequences. To obtain the probability of generating a specific sequence, one must sum the expression in Eq.\ \ref{eq:pgen} over all the recombination scenarios that result in that sequence.

We determine the component probability distributions in  Eq.\ \ref{eq:pgen}, $P({\rm V})$, $P({\rm D,J})$, $P({\rm insVD},{\rm insDJ})$, etc., directly from the data using the principle of maximum likelihood. The likelihood of a whole data set is given by the product, {over all the unique out-of-frame sequences in the data set}, of the generation probabilities of those sequences  according to the model. We restrict the dataset to out-of-frame sequences because, unlike most in-frame sequences, those sequences are not subject to any selection effect. Out-of-frame sequences only owe their survival to the creation of a productive gene in a second recombination attempt on the other chromosome, and thus give direct insight into the raw recombination process without confounding factors. In practice, likelihood maximization is performed using an Expectation-Maximization algorithm, as explained in \cite{Murugan2012}.
A distinct set of generation parameters was inferred from each of our datasets using the procedure above. These datasets included TRB sequence repertoires from the thymus, spleen and blood of mice at ages 17 days after conception, and 4 days, 21 days and 42 days after birth (see Materials and Methods and SI Table 1 for a complete summary of data).

The main assumption underlying Eq.\ \ref{eq:pgen} is its simple product structure, reflecting the independence of the enzymes that carry out different steps of the process. Another assumption is that the probability of inserting a given N nucleotide only depends on the identity of the nucleotide that precedes it (Markov assumption).  We self-consistently checked the validity of these assumptions by verifying {\em a posteriori} that almost no unaccounted correlations between the recombination events were left in the data that were not explicitly assumed (see  SI text and Figs.\ S1), and by showing that the statistics of triplets of N insertions were well predicted by the Markov model (Fig.\ S2). We also compared our probabilistically inferred distributions of recombination scenario variables with distributions assembled from assignments made by a standard VDJ alignment software package \cite{Brochet2008}. We found that these non-probabilistic alignment methods greatly overestimate
the fraction of sequences with no N nucleotide insertions, and significantly violate the D-J pairing rule imposed by genome topology, while the probabilistic method does not (Fig.\ S3). This discrepancy is what motivates our use of a probabilistic approach (see SI text for more discussion). The inferred model features were very reproducible across individuals of the same age (Fig.\ S5 and S7).

\subsection{Maturation of diversity in the thymus}
The best documented element of VDJ recombination known to change between fetal and adult life is the number of non-templated (N) insertions at the junctions. In Fig.\ \ref{fig2} we plot the marginal distributions of the number of N insertions at the VD and DJ junctions inferred from out-of-frame thymic sequences of mice at a sequence of ages. During the passage from fetal to mature animal, this distribution changes dramatically and rapidly. In the embryonic mouse, 90\% of the sequences have no insertions, while this fraction drops to 10\% in adult mice.
This trend is consistent with previous observations in neonates \cite{Feeney1990}, and is explained by the low level of expression of TdT before birth \cite{Bogue1992,Gilfillan1993a}.
TdT is turned on after birth, and Fig.\ \ref{fig2} indicates that the asymptotic level of TdT in adults must be reached before 21 days, as there are no noticeable differences between 21 and 42 days.
The distribution at 4 days shows an intermediate situation, roughly halfway between embryonic and adult.
While Fig.\ \ref{fig2} shows that the inferred distribution of insertions is identical at the VD and DJ junctions in the thymus of fetal or adult mice, it is different at 4 days. This difference could be due to the temporal ordering of VDJ recombination:
DJ recombination occurs before VD recombination, with a short time delay between the two,
and the rise of TdT expression during this delay could explain the increased mean number of VD insertions relative to that of DJ insertions.

\paneltwo 

We asked whether features of the recombination process other than the number of insertions changed with development age.
D and J gene choice did not change significantly ($P>0.05$, t-test corrected for multiple testing), while a few V genes saw their usage significantly ($P<0.05$) increase (V4, V12-1, V26) or decrease (V14, V16, V17, V20, V22) with age (SI text and Figs.\ S3--S5).
The profiles of deletion showed no significant changes with age (Figs.\ S6 and S7), and neither did the frequencies of inserted N nucleotides (Fig.\ S2).

\panelthree

To quantify the overall change in diversity between TRB sequence repertoires at different ages, we calculated the Shannon entropy of their distributions.
Entropy is an additive quantity, meaning that it can be decomposed as a sum of contributions from 
gene choices, deletions, and insertions, from which a correction for convergent recombination must be subtracted \cite{Murugan2012}.
We find that the diversity of generated nucleotide sequences increased from $21$ bits in fetal mice to $30$ bits in adult mice.
The change in repertoire diversity during this transition is almost entirely due to the change in the insertion profile, as can be seen in Fig.\ \ref{fig3} (inset) where the different contributions to the sequence entropies of the fetal and mature sequence repertoires are compared.

The entropy is mathematically equal to the negative of the mean of the logarithm of generation probability. For adult mice, Fig.\ \ref{fig3} shows that this number corresponds to a generation probability of roughly one per billion for typical sequences. The same curve also shows that the generation probabilities of individual sequences (in typical repertoires of $10^4$ to $10^5$ sequences) range from a few parts in a million to less than a part in $10^{18}$. T cells with very low generation probabilities will be private, {\em i.e.} not likely to be generated independently in two mice, while T cells with the highest generation probabilities can be public, {\em i.e.} frequently found in different mice \cite{Venturi2008a}. These numbers apply to the whole TRB nucleotide sequences. If we restrict our attention to the amino-acid sequence of the CDR3, codon degeneracy means that sequence entropies will be smaller, with typical individual sequence generation probabilities of $10^{-6}$, and a maximum of $3\cdot 10^{-4}$ (as estimated by generating $10^8$ random sequences from the model).

\subsection{The peripheral repertoire reflects past states of the thymus}
Our analysis of thymic repertoires has shown that the generative probability distribution for VDJ recombination changes dramatically in the days and weeks after birth.
In order to understand how the evolution of VDJ recombination
impacts repertoires, we need to account for the fact that
various compartments accumulate cells generated
across earlier times, as sketched in Fig.\ \ref{fig4}. As a result,
sequence repertoires must be described by a {mixture} of generative models, with varying parameters that reflect the past states of the generation process. 

\panelfour

In order to use our inference procedure to quantify the state of mixing of repertoires, we must make some simplifying assumptions. First, given our observation that other features vary rather little with age (Figs.\ S4--S9), we assume that only the statistics of the untemplated insertions change with time.
We further assume that the insertion distribution function interpolates linearly between the embryonic and adult distributions:
denoting by $n$ the number of insertions at a junction, its distribution is $P_{\alpha}(n) = \alpha P_{\rm emb}(n)+(1-\alpha)P_{\rm mat}(n)$, where 
$P_{\rm emb}$ is the distribution for the 17 day embryo, and $P_{\rm mat}$ is the adult distribution at 42 days and $0\leq\alpha\leq 1$ is an effective level of TdT measured by its impact on the number of insertions. We verified the validity of this interpolation by checking that it describes the data at day 4, shown in purple in Fig.\ 2, accurately (Kullback-Leibler divergence of $0.1$ bits, which is small compared to the entropy of $2.6$ bits).

The distribution $P_{\alpha}$ describes the TRB generation process at a fixed TdT level $\alpha$. 
As explained above, repertoires represent the accumulated output of recombination events at earlier times and must be described by a mixture of processes at various $\alpha$ values. The generic mixture model for insertions $n_1, n_2$ at the VD and DJ junctions can thus be written as
\beq
\begin{split}
{P}(n_1, n_2) &= \int_0^1 d\alpha\, g(\alpha) {P}_\alpha (n_1) {P}_\alpha(n_2) \\
&=   {P}_{\bar\alpha}(n_1) {P}_{\bar\alpha}(n_2) + {\rm var}(\alpha) \Delta {P}(n_1) \Delta {P}(n_2),
\end{split}
\label{eq:mixmod}
\eeq
where $g(\alpha)$ is the distribution of $\alpha$ in the repertoire reflecting the distribution of the past developmental ages at which its receptors were produced, 
$\bar\alpha$ and ${\rm var}(\alpha)$ are its mean and variance, and $\Delta P= P_{\rm mat} - P_{\rm emb}$. Conveniently, per the second line of Eq.\ \ref{eq:mixmod}, the mixture distribution depends only on the mean and variance of $\alpha$.
The variance is constrained by $0 \leq {\rm var}(\alpha)\leq \bar\alpha(1-\bar\alpha)$, and gives a measure of the level of mixing in the repertoire. Zero variance means no mixing, {\em i.e.} all cells were created at a single effective TdT level $\alpha=\bar\alpha$. Maximal variance and mixing are attained when a fraction $\bar\alpha$ of cells fully expresses TdT ($\alpha=1$), while the remaing fraction $1-\bar\alpha$ does not express TdT at all ($\alpha=0$).

We estimate $\bar \alpha$ and ${\rm var}(\alpha)$ for our datasets by first inferring the joint distribution $P({\rm insVD,insDJ})$ of Eq.\ \ref{eq:pgen} using our inference technique, and then adjusting $\bar \alpha$ and ${\rm var}(\alpha)$ to obtain the best fit to Eq.\ \ref{eq:mixmod}.
The data points in Fig.\ \ref{fig5}A show the mean effective TdT level $\bar\alpha$ as a function of age for thymus or blood data sets,
while the data points in Fig.\ \ref{fig5}B report the associated values of var($\alpha$). Fig.\ \ref{fig5}A shows that the blood repertoire transitions from embryonic ($\alpha=0$) to mature ($\alpha=1$) with a time delay relative to the thymic repertoire. This result is expected since blood T cells are first produced in the thymus. The rise in TdT level results in an increase of diversity, as measured by the entropy of recombination events (inset). 
Fig.\ \ref{fig5}B  shows that, while embryonic and adult repertoires have no mixing, ${\rm var}(\alpha)\approx 0$, all intermediate repertoires are mixed, with ${\rm var}(\alpha)$ significantly larger than 0. Although the thymus does not accumulate cells, T cells do spend a finite time in the thymus and, when TdT levels are changing fast, thymic repertoires are described by mixtures. Still, blood repertoires are substantially more mixed than thymic repertoires, as is expected since the thymus contains cells that have recombined over a narrow range of TdT levels $\alpha$, while blood contains cells with a greater range of ages and of values of $\alpha$.

\panelfive

The behavior of the data displayed in Fig.\ \ref{fig5} can be better understood by comparison with a simple model (see Materials and Methods and SI). In this model  the effective TdT level $\alpha(t)$ in the thymus is given by a sharply rising Hill function (Fig.\ \ref{fig5}A, dashed curve), recombined cells are created at a rate that increases rapidly with time, and cells reside in the thymus for three days on average, after which they are released into the periphery. While model parameters were chosen to reproduce the observed behaviour quantitatively, we did not attempt a formal fit to the data, because of the paucity of data points. Results for $\bar\alpha(t)$ and ${\rm var}(\alpha)(t)$ are displayed in Fig.\ \ref{fig5} (orange and green curves). The model recapitulates the delay in maturation between thymus and blood (Fig.\ \ref{fig5}A) and also accounts for the observed level of mixing as a function of time in blood and thymus (Fig.\ \ref{fig5}B). The model curves in Fig.\ \ref{fig5}B are parametric in time (time stamps added for clarity) and it is significant that the data points lie close to points on the model curves at the right age.

\subsection{Selection}

Our discussion so far has been restricted to the evolution of the generative model for VDJ recombination, which is inferred from nonproductive, out-of-frame sequences. The statistics of in-frame sequences, which code for functional surface receptor proteins, is biased by the selection effects they have been subjected to, both in thymus and periphery. These selection effects can be quantified from the statistical differences between the selected, inframe repertoire and the unselected, out-of-frame repertoire.

Following previous work on human TRB sequences \cite{Elhanati2014}, we associate to each possible CDR3 amino-acid sequence of the TRB repertoire a selection factor $Q$ defined as the fold-change in probability between generation (inferred from out-of-frame sequences) and the observed statistics of unique in-frame sequences. The selection factor $Q$ is assumed to be a product of sub-factors related to 
CDR3 length $L$, $q_L$, and amino-acid identity $a$ at each position $i$ of the CDR3, $q_{i,L}(a)$. 
The set of these subfactors defines a motif of selection across all possible TRB sequences.

We inferred selection motifs for thymic and blood repertoires by maximizing the model likelihood over the selection subfactors (see Materials and Methods and Fig.\ S8). The composition selection factors $q_{i,L}(a)$ were very consistent between thymus and blood of mice of the same age, with weaker consistency between mice of different ages (Fig.\,\ref{fig6} and Fig.\ S9). Similarity between blood and thymus may seem surprising, as we could have expected a significant fraction of TRB from thymic cells to have been sequenced before any selection effect, making them statistically closer to out-of-frame sequences. These observations suggest that our selection factors primarily capture selection for the ability of the coded protein to fold into a displayable receptor, and may not capture more subtle effects such as negative selection against self-recognition. We report the patterns of selection in terms of amino-acid biochemical properties in Fig.\ S8.

\panelsix


\section{Discussion}
VDJ recombination is a highly stochastic process that produces the initial diversity on which the adaptive immune system relies to develop a functional and diverse repertoire of receptor specificities. Previous studies have shown that this diversity is limited in neonates compared to adults, either by biasing the choice of gene segments \cite{Yancopoulos1984,Perlmutter1985,Schroeder1988,Jiang:2011dt}, or by having a small number of N insertions \cite{Feeney1990,Feeney1991}. Combining high-throughput sequencing with statistical analysis of murine T-cell receptor beta chains, we analyzed the dynamics of maturation of VDJ recombination. This analysis allowed us to precisely quantify, in bits, how diversity increases with age, from embryo to adult. We found that the most significant change in the recombination statistics was the number of untemplated N insertions, which sharply increases around the age of 4 days, from almost no insertions to the amount found in adults. Low number of insertions in neonates and during embryonic development are common to both B- and T cell receptors \cite{Schroeder2001}, and is attributed to low TdT expression \cite{Bogue1992,Komori1993a,Gilfillan1993a}. Diversity can be further reduced in embryo by
concentrating gene usage on only a few combinations, as was shown for immunoglobulin in mouse \cite{Yancopoulos1984,Perlmutter1985}, human \cite{Schroeder1988}, and more recently zebrafish using high-throughput sequencing \cite{Jiang:2011dt}. Similar observations were made on human TCR beta chains \cite{George1992,Raaphorst1994}. By contrast, we found only minor differences in TRB gene usage between embryonic and adult mice (Fig.\,S6), meaning that the reduced number of N insertions is the only factor limiting diversity in the embryo relative to adult.

One can only speculate about the biological function of the lack of N insertions in embryos and very young individuals. Rearrangements with no insertions may encode particular specificities that are effectively innate. An example of such genetically coded receptors are the invariant TCR of MAIT and NKT cells \cite{LeBourhis2011}. These TCRs, which lack N insertions, are formed with high probability by VDJ recombination \cite{Greenaway2013}, and are further selected to be very conserved. Receptors lacking N insertions may provide neonates with a minimal set of innate-like specificities, ensuring basic immunity \cite{Gilfillan1995}, which is later completed by the full diversity of receptors endowed with N insertions. 

Our analysis highlights the importance of focusing on the underlying statistical ensembles from which repertoires are drawn, rather than looking for significance in the sequences themselves. While sequence repertoires are contingent and noisy, with little to no overlap between individuals, sequence ensembles are consistent between individuals, as was already noted for humans TCR \cite{Murugan2012,Elhanati2014}. Crucially, a statistical treatment is essential for tracking the precise dynamics of N insertions with age, as deterministic assignments give systematically biased estimates of these numbers (Fig.\,S3).
When studying the development of repertoire diversity, another confounding factor is the selection of particular receptor features during development, notably through thymic selection. This selection process could also change with developmental age. By focusing our analysis on non-productive rearrangements, we deconvolved these selection effects from the intrinsic changes in the recombination machinery. Analysing the productive repertoire, we found no significant differences in selection between embryonic and adult receptors.

The study of mouse raises interesting and puzzling questions about sequence diversity. We found that the mature mouse repertoire is 9 bits (or $2^9\approx 500$-fold) more diverse than the embryonic repertoire.
On the other hand, the mature mouse repertoire is 14 bits (or $2^{14}\approx 16,000$-fold) less diverse than the human T cell repertoire, owing to a lower number of N insertions (typically 2 per junction in mouse versus 5 in human). Human and mouse have to deal with presumably equally complex pathogen environments, and it would be natural to expect their immune systems to have similar levels of sequence diversity. However, this $\sim 10,000$-fold difference in potential diversity closely reflects the difference in the number of T cells in the two species, $\sim 10^7$ in mice {\em vs.} $\sim 10^{11}$ in humans.
Another difference with humans is the timing of the transition. The number of  TCR N insertions increases as early as the first semester of gestation in humans \cite{George1992}, and from the second semester for BCRs \cite{Schroeder2001}. By contrast, our results for mice show a sharp transition soon {after} birth.

The generative models of TCR inferred as part of this study could be used to investigate the existence of a public repertoire. We estimated the typical generation probability of TRB nucleotide sequences in adult animals to be $\sim 10^{-9}$ (Fig.~3). Since mice have $\sim 10^7$ cells, this suggests that most of their nucleotide sequences are private. However, not all sequences are equally likely, and some have generation probabilities as high as $\sim 10^{-5}$, meaning that they may be expected to be found in all individuals, and provide the basis for the public repertoire. This effect is even more marked when considering CDR3 amino-acid sequences, which have typical generation probabilities of $10^{-6}$, suggesting that a significant fraction of them may be public. Interestingly, since diversity is limited in neonates, the typical receptor generation probabilities are much higher, which further supports the hypothesis that many of these low diversity TCRs are public and innate-like.

\section{Materials and Methods}

\subsection{Datasets}
The data used in our analyses are 87bp (and 60bp) nucleotide sequences covering the variable region of the rearranged mouse TRB gene. The sequences were obtained by Adaptive Biosciences using their TRB DNA sequencing protocol (including error correction on the basis of multiple reads of each unique DNA sequence) applied to biological samples provided by two of the authors (AL and CD). The samples comprised blood, spleen and thymus samples from mice sacrificed at four different ages: 17 day embryo, 4 days, 21 days, and 42 days post birth (the library preparation and sequencing for day 42 thymic samples were replicated). The mice were Black 6 laboratory mice (Jackson Labs) raised in standard laboratory conditions. 
The number of unique sequences in the various data sets were a few tens of thousands on average (with a few data sets providing more than $10^5$ unique sequences). The sequencing of the mature (D42) thymus samples was replicated once.
Detailed statistics on the data sets are provided in SI Table 1. The full sequence data sets are available, along with an explanatory README file, at \url{http://princeton.edu/~ccallan/MousePaper/data/}. Unless otherwise stated, the presented statistics are based on aggregates of different mice.

\subsection{Model inference}
The sequence data are used as inputs to computational statistical inference procedures designed to construct the probability distributions on sequence space from which the individual data sets may be seen as independent draws. Two distinct inference procedures are used. The first uses out-of-frame data sequences to infer the statistical ensemble of sequences produced directly by the VDJ recombination process. This procedure is described in detail in \cite{Murugan2012}. The second uses in-frame data sequences to infer the modifications to the initial statistical ensemble that are brought about by selection of recombined TRB genes for proper immunological function. This second procedure is described in detail in \cite{Elhanati2014}. The Matlab software for implementing the two procedures is available at \url{http://princeton.edu/~ccallan/MousePaper/software/}. The results of the inference procedure (in the form of Matlab data files), along with instructions on how to use these files to recreate the figures in this paper are available at \url{http://princeton.edu/~ccallan/MousePaper/results/}. 

\subsection{Model of repertoire maturation}
TCR are produced in the thymus with a time-dependent effective TdT level $\alpha(t)=[1+((T_{\rm half}-T_{\rm start})/(t-T_{\rm start}))^{20}]^{-1}$, with a production rate $\theta(t) \propto (1 + (t - T_{start} + 1)^{2.1})$ (arbitrary units). Time is in days, with birth at $t=0$, $T_{\rm start}=-15$ and $T_{\rm half}=2$. Cells reside in the thymus for an average of 3 days (exponentially distributed time), after which they are released into the periphery. The simulation is followed from $t=T_{\rm start}$ (early embryo) to $t=42$ (age of oldest data set).

{\bf Acknowledgements.} The work of YE, QM, TM and AW was supported in
part by grant ERCStG n. 306312. The work of CC and ZS was supported in
part by NSF grant PHY-1305525.
\medskip

\appendix

\setcounter{figure}{0}
\makeatletter 
\renewcommand{\thefigure}{S\@arabic\c@figure}
\makeatother

\graphicspath{{"./SuppInfo/"}}
\makeatletter
\def\input@path{{"./SuppInfo/"}}
\makeatother

\section*{Supporting Information}

\section{Validation of the structure of the sequence generation model}
Our inference procedure rests on a presumption of independence of the various factors in the generative model and a verification of that independence is an important aspect of our analysis. To address this issue, we compute the mutual information -- a non-parametric measure of dependence between random variables -- between the various recombination scenario variables, and compare these numbers between the data and the generative model inferred from the same data.

The mutual information between two random variables $x$ and $y$ jointly distributed according to $p(x,y)$ is defined as:
\beq
I(x,y)=\sum_{x,y}p(x,y)\log\frac{p(x,y)}{p(x)p(y)}.
\eeq
The mutual information between the variables defining recombination scenarios, as computed from the inferred model itself, can be calculated exactly and are shown in the below-diagonal halves of the matrices of Fig.\ S1. By construction, the generative model has zero mutual information between certain variable pairs, {\em e.g.} the number of VD insertions and the choice of J gene, and non-zero mutual information between variables that correlate with each other either directly or indirectly, {\em e.g.} between D and J gene choice, or between $D$ choice and number of $D$ deletions.

On the other hand, the inference procedure assigns multiple scenarios, each with its own probability, to each data sequence. For any pair of scenario variables ({\em e.g.} insVD and delJ) one can use these assignments over all the data sequences to populate a list of pairs of values, weighted by scenario probabilities. From this list we can then compute the mutual information between the two elements of the pair using the Treves-Panzeri correction ({\em Neural Computation}, {\bf 7}(2) 399, 1995) to account for small sample sizes. The mutual informations computed in this fashion are displayed in the above-diagonal halves of the matrices of Fig.\ S1. The model form is considered accurate if the obtained mutual informations agree with those predicted by the model, {\em i.e.} if the matrices of Fig.~S1 are symmetric.

\section{Mixture model}
In this model, T cells are introduced into the thymus with rate $\theta(t)$, with an effective TdT level $\alpha(t)$. Cells leave the thymus into the periphery (blood and spleen) with constant rate $\tau_{\rm thy}^{-1}$. Under these assumptions, the total number of cells in the thymus, $N_{\rm thy}(t)$, and the periphery, $N_{\rm peri}(t)$, read:
\begin{eqnarray}
N_{\rm thy}(t) &=& \int_{-\infty}^t \theta(t') e^{-t'/\tau_{\rm thy}} dt',\\
N_{\rm peri}(t) &=& \int_{-\infty}^t \theta(t') \left( 1 - e^{-t'/\tau_{\rm thy}} \right) dt'.
\end{eqnarray}
The distributions of $\alpha$ in these compartments are given by:
\begin{eqnarray}
g_{\rm thy}(\alpha;t)&=&\frac{1}{N_{\rm thy}(t)}\int^t \theta(t') e^{-\frac{t'}{\tau_{\rm thy}}}
\delta(\alpha(t')-\alpha) dt',\\
g_{\rm peri}(\alpha;t)&=&\frac{1}{N_{\rm peri}(t)}\int^t \theta(t') (1- e^{-\frac{t'}{\tau_{\rm thy}}})
\delta(\alpha(t')-\alpha) dt'. \nonumber
\end{eqnarray}

The mean and variance of $\alpha$, plotted in Fig.~5 of the main text, are calculated from these expressions. They are sufficient to calculate the joint distribution of insertions at the two junctions, per Eq.~2 of the main text.

\section{Selection model}
The selection factor is defined as the fold-change between the probability of generation of a TRB sequence $s$, $P_{\rm gen}(s)$, and its probability among productive in-frame sequences, $P_{\rm post}(s)$:
\beq
P_{\rm post}(s)=Q(s)P_{\rm gen}(s).
\eeq
We assume that $Q(s)$ depends on $s$ only through the amino-acid translation of its CDR3, and that it takes the following factorized form:
\beq
Q(a)=q_L \prod_{i=1} q_{i;L}(a_i),
\eeq
where $a=(a_1,\ldots,a_L)$ is the amino-acid sequence of the CDR3, and $L$ its length. The $q_L$ factors are length-specific factors, while $q_{i;L}(a)$ are composition-specific factors. All factors are simulatenously inferred by maximizing the likelihood of the inframe sequences using gradient ascent, as explained in detail in Elhanati {\em et al.}, {\em PNAS} {\bf 111}:9875 (2014).
Fig.~S8 shows the values of $q_{i;L}(a)$ inferred from D42 mouse thymic sequences, while Fig.~S9 compares the values of $q_{i;L}(a)$ inferred from different datasets. The selection factors were normalized for these figures such that $q_{i;L}(a)>1$ indicates a positive contribution to the overall selection, while a value below $1$ indicates a negative contribution.
A different and model-independent normalization scheme was used for the comparisons in figure 6 in the main text, such that $\sum_a q_{i;L}(a)=1$.

\bibliographystyle{pnas}

\newpage

\begin{table*}[htbp]
  \centering
  \caption{{\bf Sequence data sets used in analyses reported in main text.}}
    \begin{tabular}{ccccc}
    \\
    Age  & Tissue & Read length & Productive & Nonproductive \\
    \hline
    17 day embryo & Thymus 2 & 87 bp & 11199 & 15368 \\
    17 day embryo & Thymus 3 & 87 bp & 9589  & 11485 \\
    4 day & Thymus 1 & 87 bp & 9165  & 4687 \\
    4 day & Thymus 2 & 87 bp & 27794 & 14417 \\
    4 day & Spleen 1 & 87 bp & 447  &  325\\
    4 day & Spleen 2 & 87 bp & 213 & 112 \\
    4 day & Spleen 3 & 87 bp & 1356 & 668 \\
    21 day & Thymus  & 60 bp & 95164 & 43418 \\
    21 day & Blood & 60 bp & 14469 & 7510 \\
    42 day & Thymus 1 & 87 bp & 33028 & 16159 \\
    42 day & Thymus 2 & 87 bp & 24292 & 10864 \\
    42 day & Thymus 3 & 87 bp & 24846 & 12006 \\
    42 day & Thymus 1 library replicate & 87 bp & 137233 & 67232 \\
    42 day & Thymus 2 library replicate & 87 bp & 61990 & 30390 \\
    42 day & Thymus 3 library replicate & 87 bp & 83591 & 40425 \\
    42 day & Blood 1 & 87 bp & 16642 & 7550 \\
    42 day & Blood 2 & 87 bp & 3256  & 1235 \\
    42 day & Blood 3 & 87 bp & 16858 & 7306 \\
    \end{tabular}
  \label{tab:datasummary}
\end{table*}

\newpage

\begin{figure*}
\begin{center}
\includegraphics[width=\linewidth]{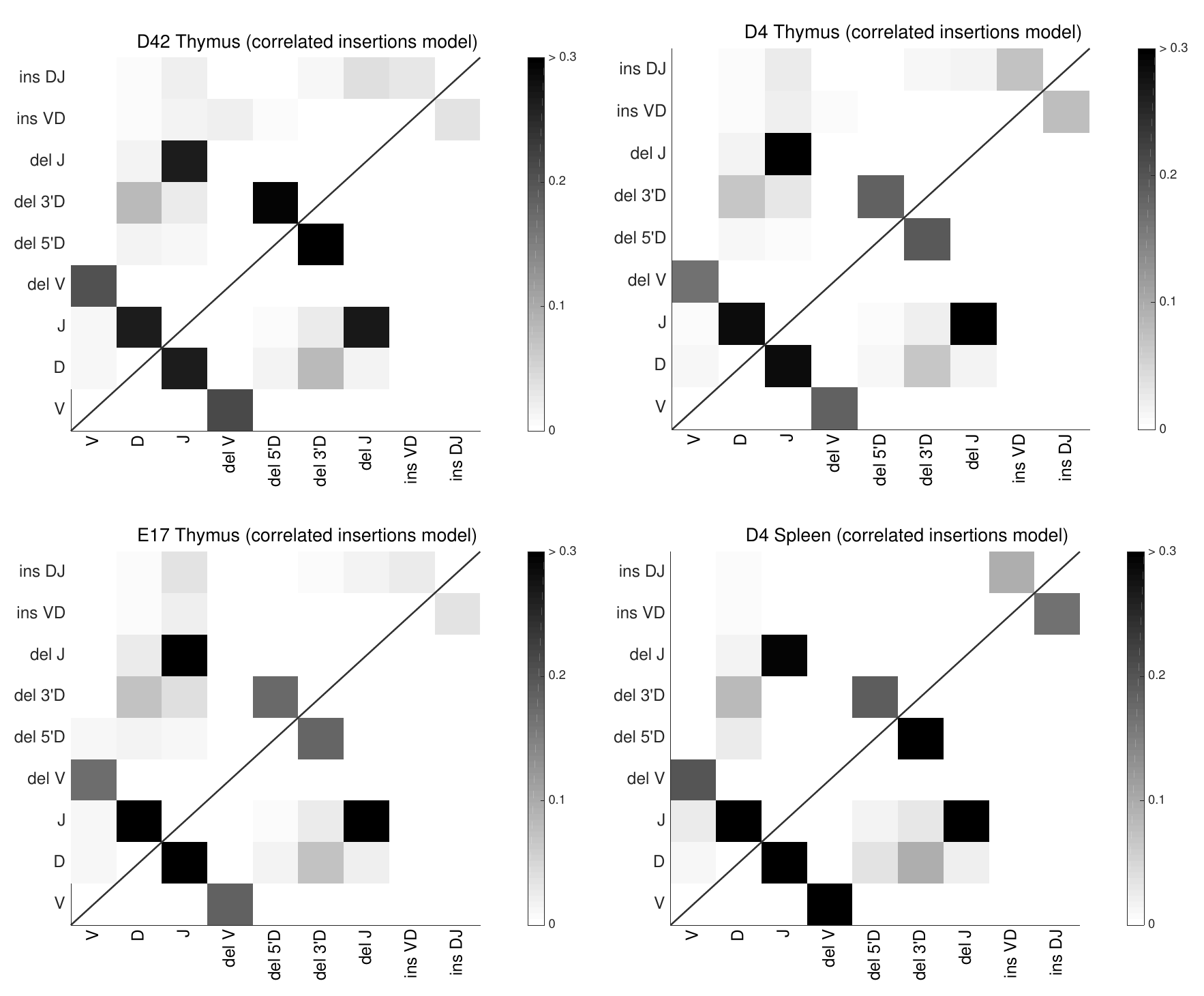}
\caption{
Mutual information (MI) between hidden scenario variables. Each plot is based on the data set indicated in the plot title and the \lq correlated insertions\rq\, model inferred from that data set. The \lq correlated insertions\rq\, models allow a general joint distribution $P({\rm insVD, ins DJ})$ of VD and DJ insertions. Squares below the diagonal (lower right) display MI values for the model itself, while squares above the diagonal display MI values derived from the data (see SI text for details). The sum of all the MI values calculated from the data (entries above the diagonal) that were not predicted by the model (zeros below the diagonal) does not exceed 0.14 bits in all considered cases. That number is very small compared to the individual entropies of each the scenario variables (each of the order of bits), indicating that these correlations are negligible.
 \label{fig:SuppZero_ter}
}
\end{center}
\end{figure*}

\begin{figure*}
\begin{center}
\includegraphics[width=.7\linewidth]{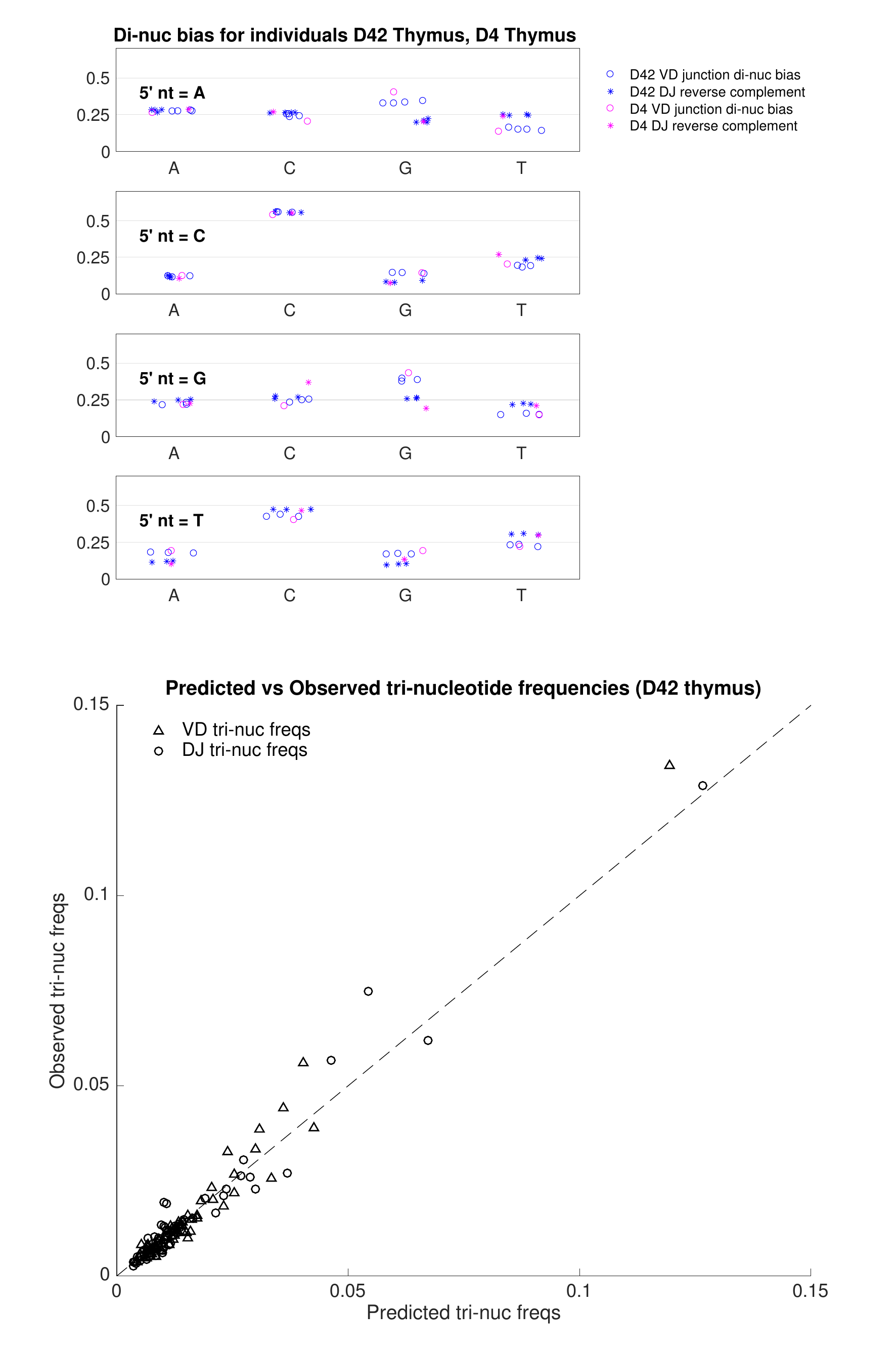}
\caption{
{\bf Frequencies of N insertions.} Upper panel: Dinucleotide
insertion bias values $P(s_i|s_{i-1})$ inferred from individual thymus
data for D4 and D42. Data are not presented for
E17 because there are virtually no insertions in embryonic
recombination. The 5' nt refers to $s_{i-1}$, while the x axis
corresponds to $s_i$.
Sample variance at a given age is small, and frequencies are
consistent from day 4 to day 42. Lower panel: predicted vs. observed
trinucleotide insertion frequencies $P(s_i,s_{i+1},s_{i+2})$ for D42 thymic data. The small scatter about the equality line indicates that the untemplated insertions are well-described by a dinucleotide Markov process. \label{fig:SuppTwo}
}
\end{center}
\end{figure*}

\begin{figure*}
\begin{center}
\includegraphics[width=\linewidth]{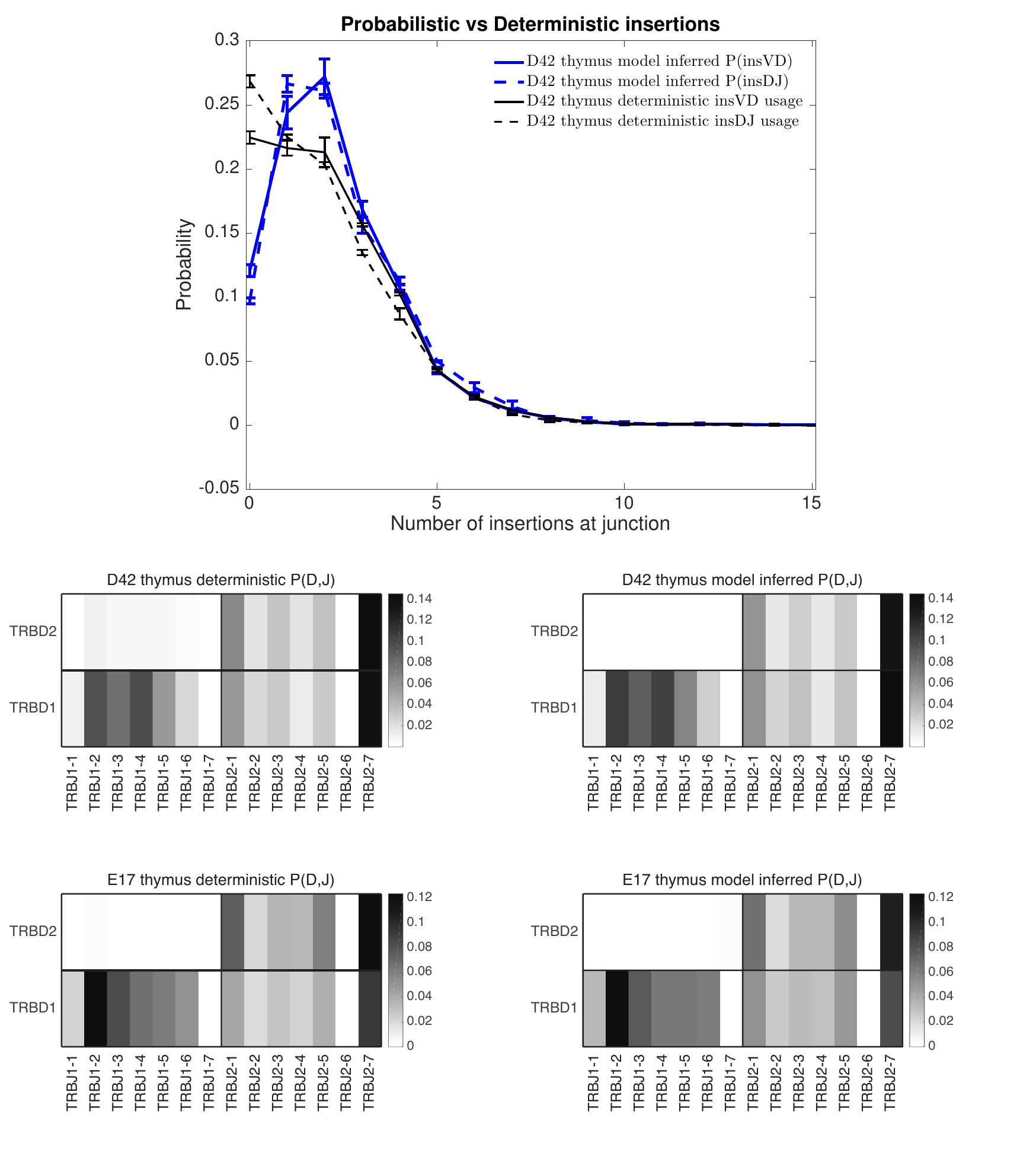}
\caption{
{\bf Comparison of deterministic vs model inference results.} We compare the results of our model inference procedure used in the paper to a deterministic alignment as described in Yousfi Monod et al, {\it Bioinformatics} 20 1:379?85 (2004). The upper panel shows the insertion profiles for D42 thymus (error bars are over individual mice). We see that the deterministic result gives a much higher probability for zero insertions and the VD and DJ junctions differ slightly. The lower panel compares the deterministic and the model-inferred joint DJ usages for D42 and E17 thymus data. Note that the deterministic alignment gives a non-zero probability of DJ pairings which are topologically impossible (i.e. TRBJ1 genes cannot be paired with TRBD2). By contrast the model-inferred usage is completely consistent with the topological constraint. \label{fig:SuppSix}
}
\end{center}
\label{fig:SuppSix}
\end{figure*}

\begin{figure*}
\begin{center}
\includegraphics[width=.8\linewidth]{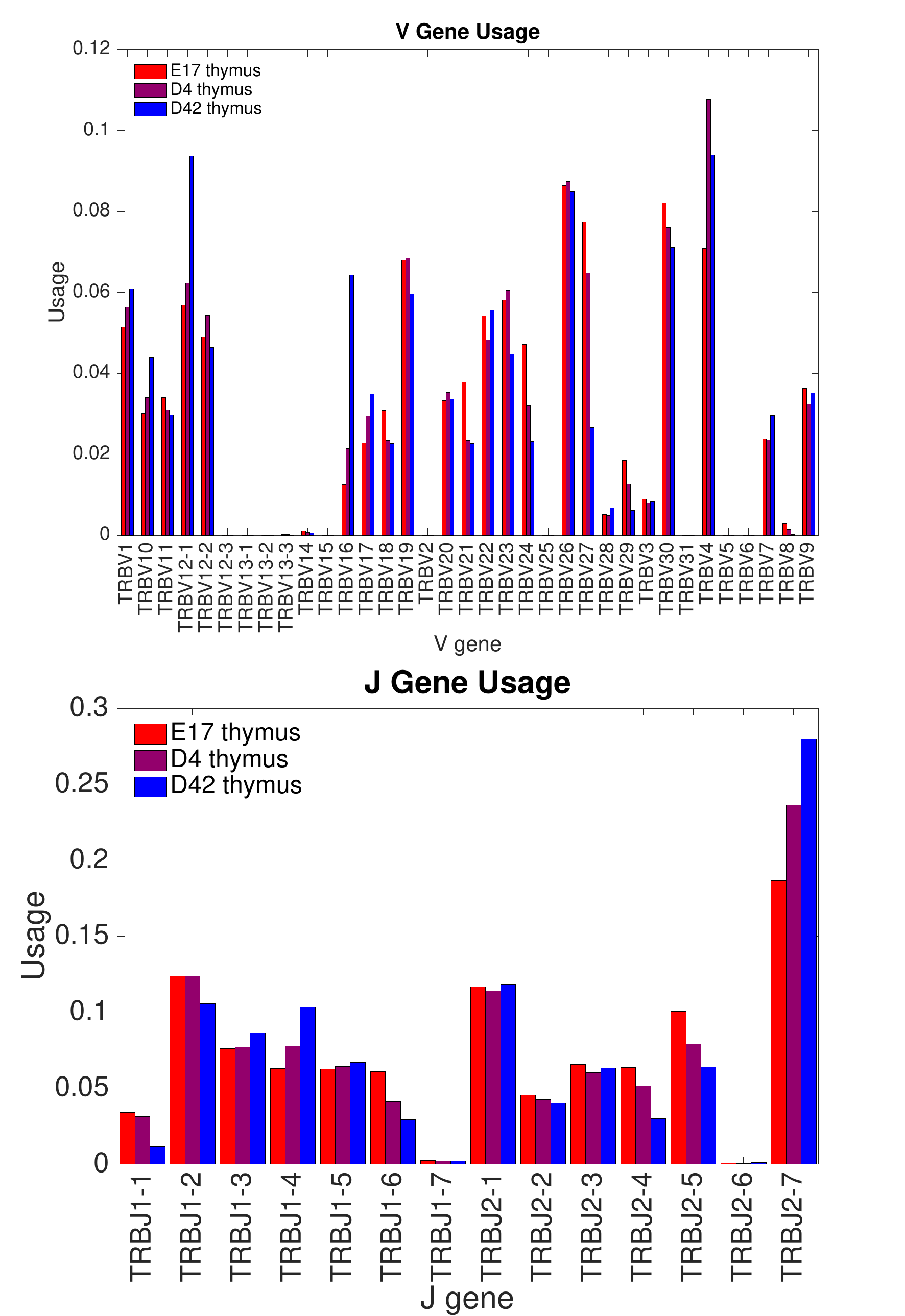}
\caption{
{\bf V and J gene usage as a function of age.} Gene usage probabilities derived from aggregated thymic data sets at different ages. Aggregated data sets, used to reduce small sample noise, are constructed by combining sequence repertoires from several individual mice at the same age. While the overall usage pattern is fairly stable with age, some genes undergo quite substantial changes in usage. \label{fig:SuppFour}
}
\end{center}
\end{figure*}

\begin{figure*}
\begin{center}
\includegraphics[width=\linewidth]{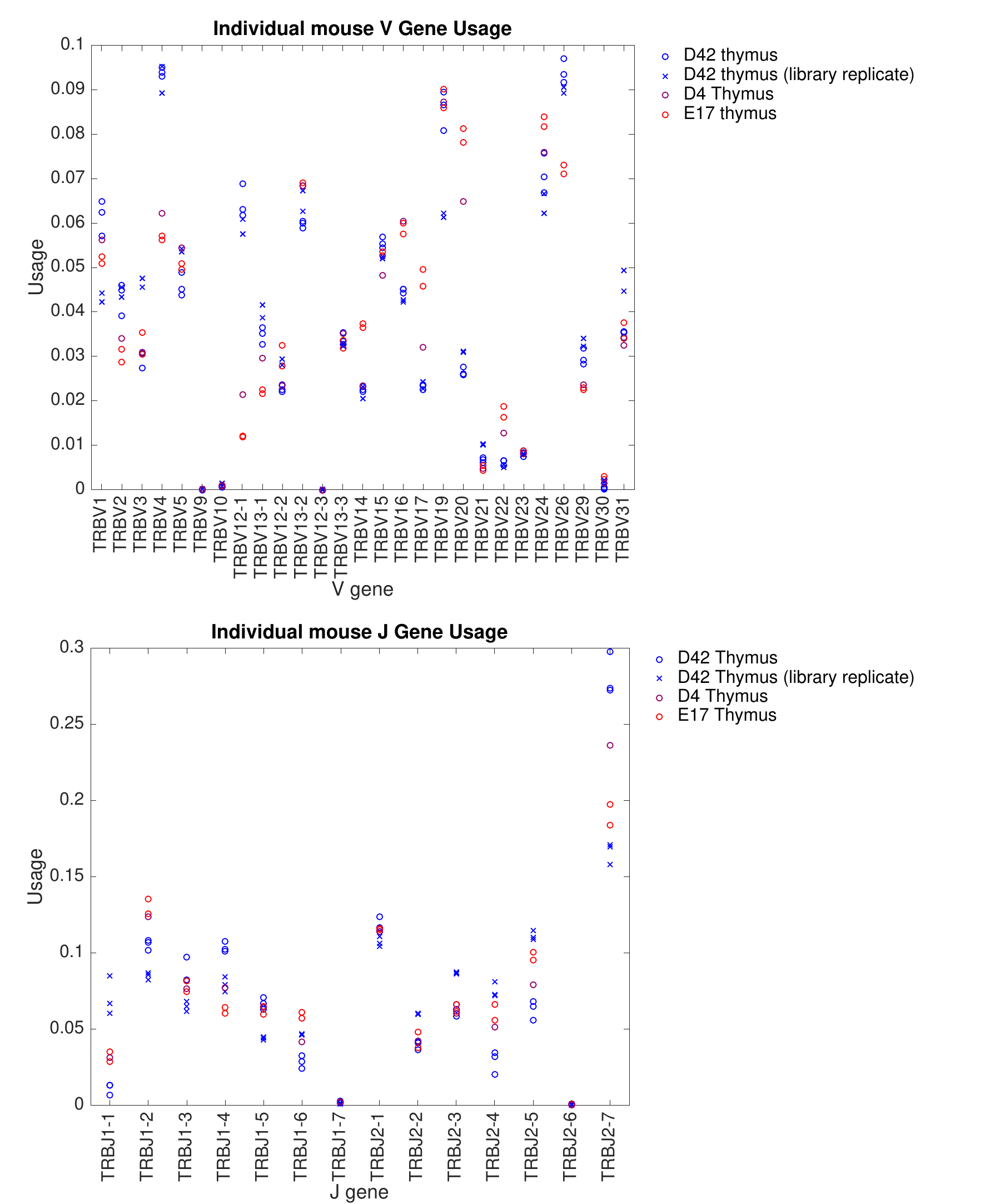}
\caption{
{\bf Variation of V and J gene usage across biological replicates.}  For each V or J gene we plot the thymic usage fraction for all available individuals (one, two or three depending on the case) for the embryonic day 17 and for the 42 day mature mouse.  Different individuals are indicated by repetitions of the same symbol, and their dispersion gives a rough measure of the biological sample variance of gene usage. \label{fig:SuppFive}
}
\end{center}

\end{figure*}

\begin{figure*}
\begin{center}
\includegraphics[width=\linewidth]{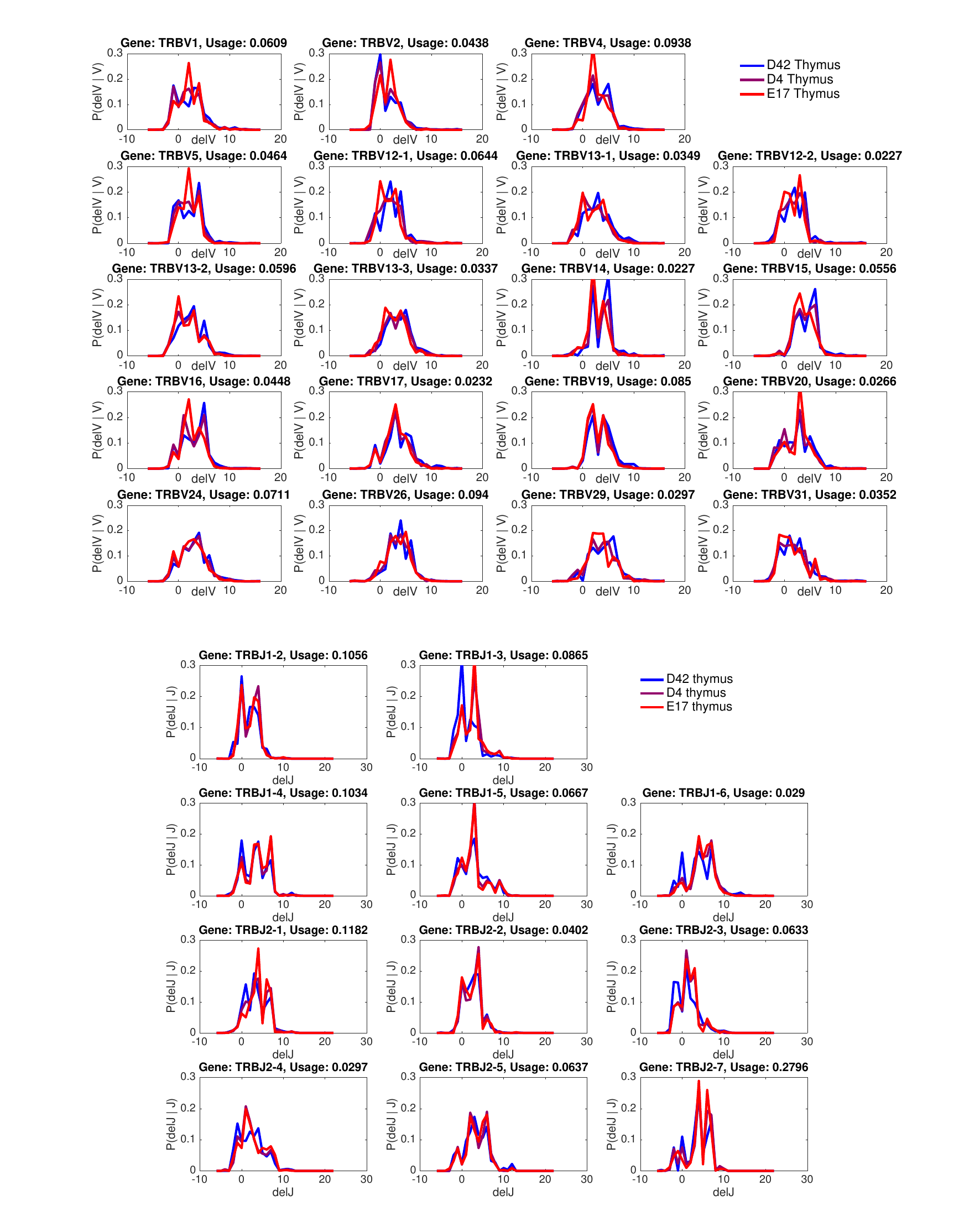}
\caption{
{\bf  Deletion profiles for different ages.}  The plots show the deletion profiles inferred from aggregated thymic data sets at three different ages. Deletion profiles depend on the identity of the gene being deleted and negative deletions are used to account for P-nucleotides. Variations across sample ages is significantly larger than individual-to-individual variations at any given age. Plot titles record the gene identities and their usage probabilities at age 42 days (only genes with usage probability greater than .02 are displayed). \label{fig:SuppThree}
}

\end{center}
\end{figure*}

\begin{figure*}
\begin{center}
\includegraphics[width=\linewidth]{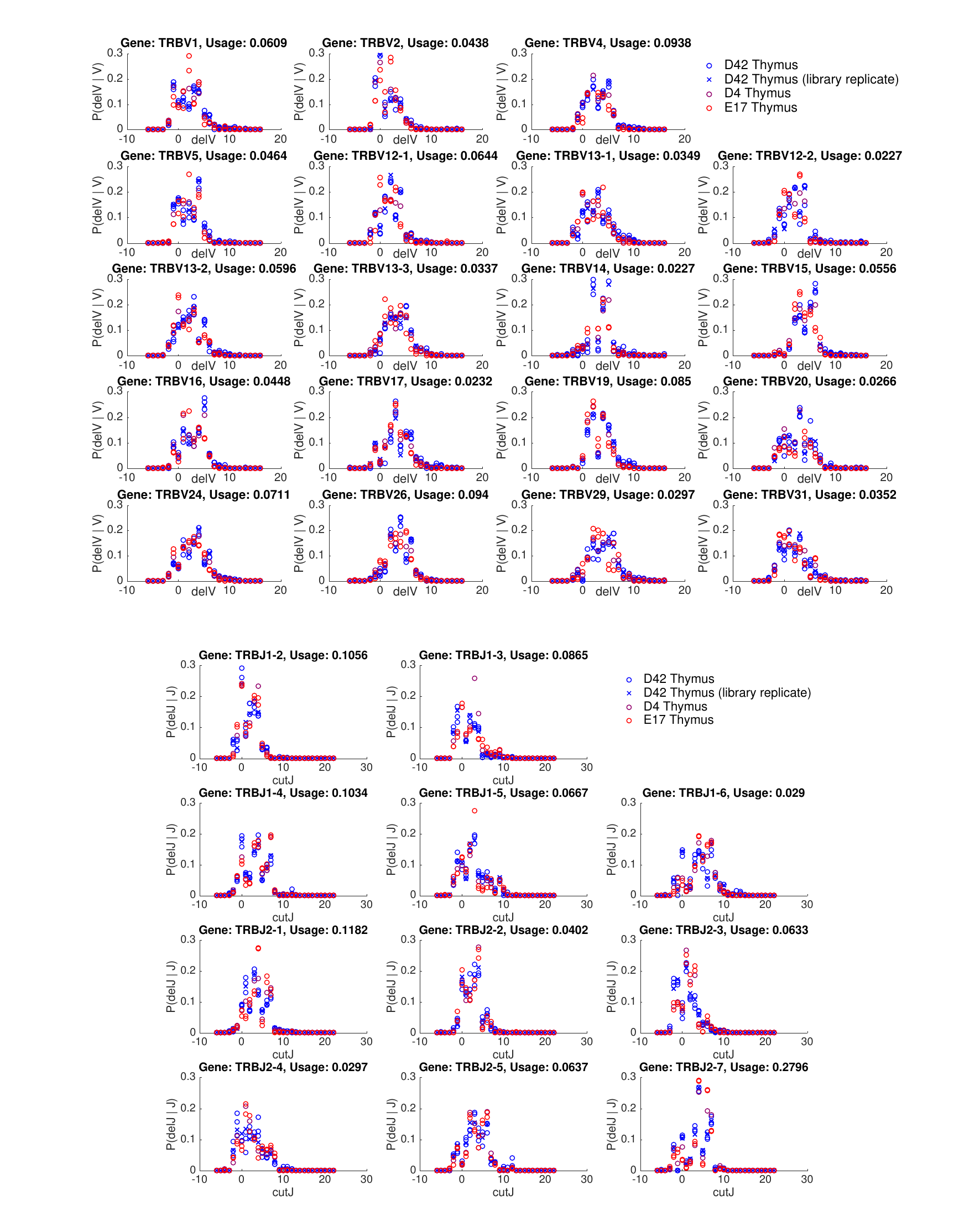}
\caption{
{\bf  Variation of deletion profiles across biological replicates.}  Same convention as Fig.~S5. Plot titles record the gene identities and their usage probabilities (only genes with usage probability greater than .02 are displayed). There is virtually no biological sample variance between individuals in a given sequencing run, and very little between runs. \label{fig:SuppThree_bis}
}

\end{center}
\end{figure*}

\begin{figure*}
\begin{center}
\includegraphics[width=\linewidth]{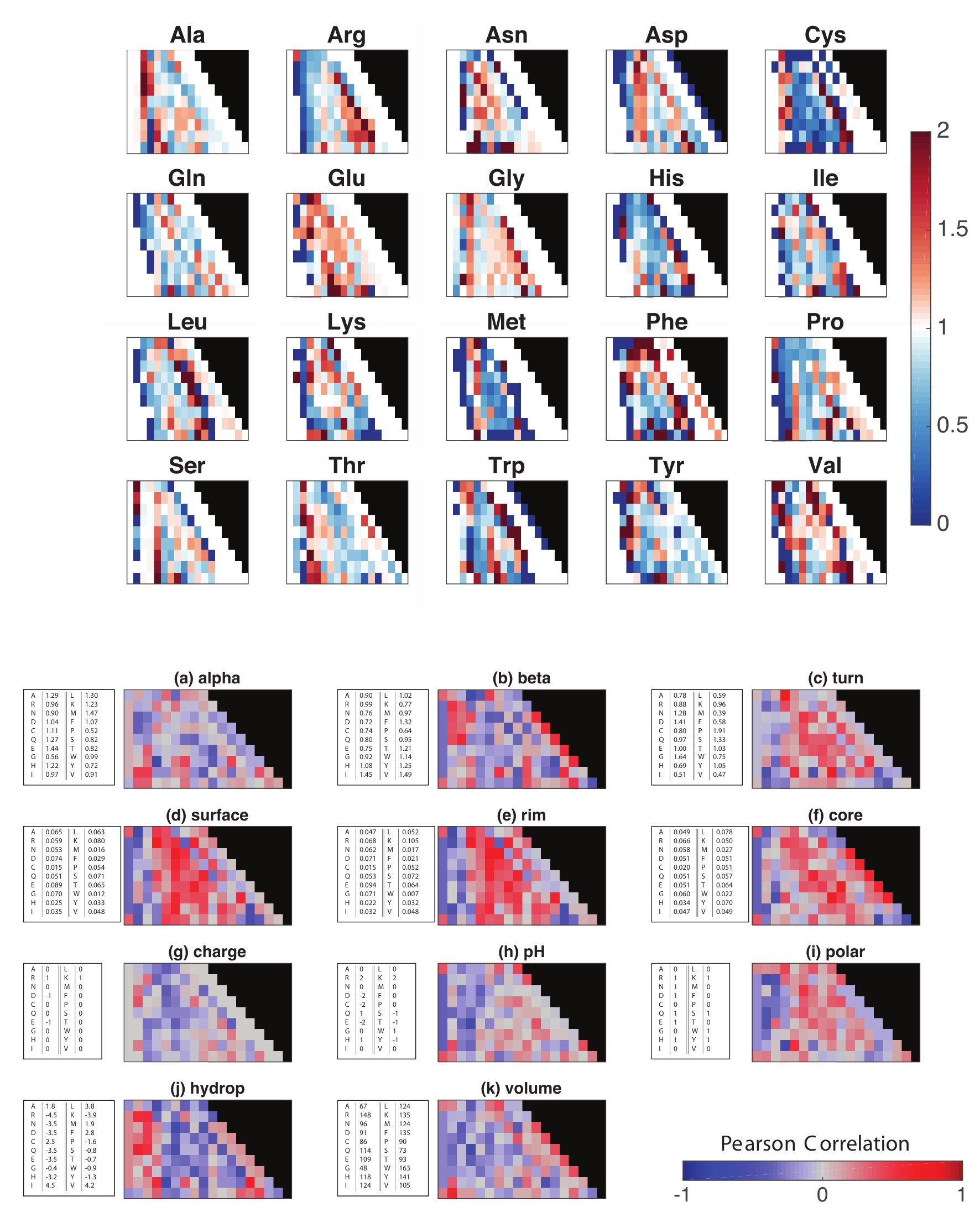}
\caption{
{\bf Selection factors.} Upper panel: Selection factors  $q_{i,L}(a)$ inferred from D42 mouse thymic data sets. Lower panel: Pearson correlations between selection factors and amino acid biochemical properties. For each position $i$, CDR3 length $L$, and biochemical property, we display the Pearson correlation between $q_{i,L}(a)$ and the values of the biochemical property for the 20 amino acids $a$ (the latter listed in the table accompanying each subpanel). \label{fig:SuppSeven}
}
\end{center}

\end{figure*}

\begin{figure*}
\begin{center}
\includegraphics[width=\linewidth]{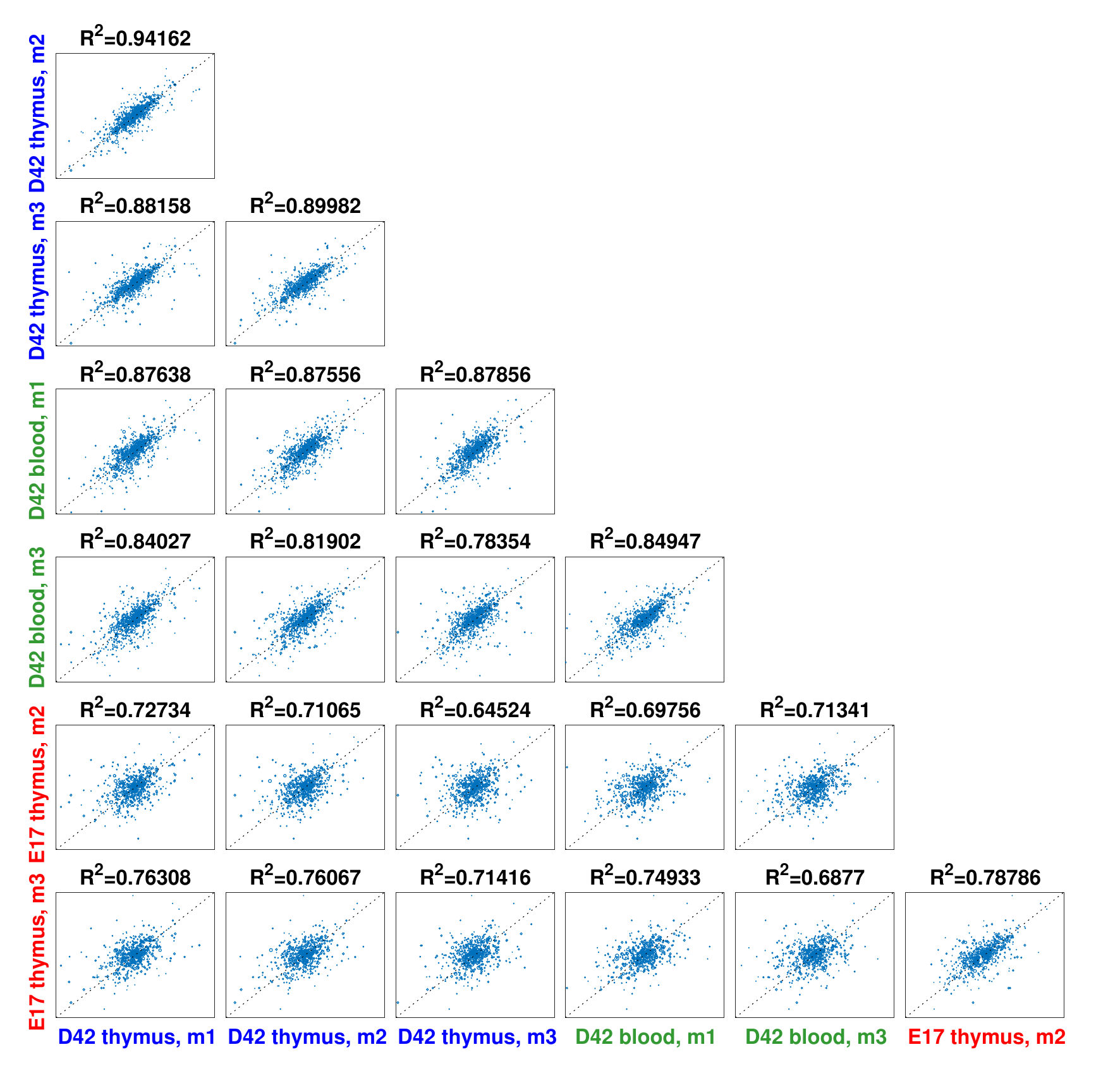}
\caption{
{\bf Selection factor correlations.} Scatter plot of selection factors inferred from pairs of data sets. The pairings allow us to compare different tissues (blood or thymus) and ages. $R^2=1-\textrm{Var}(q'-q)/(\textrm{Var}(q')+\textrm{Var}(q))$ is a measure of the difference between two data sets; for identical data sets, one would find $R^2=1$.\label{fig:SuppEight}
}
\end{center}

\end{figure*}

\end{document}